# Physical properties of single-crystalline fibers of the colossal-magnetoresistance manganite $La_{0.7}Ca_{0.3}MnO_3$

**C. A. Cardoso and F. M. Araujo-Moreira***
*Departamento de Física, Grupo de Materiais e Dispositivos - MCDCM, UFSCar, Caixa Postal 676, São Carlos SP, CEP 13565-905 Brazil*

**M. R. B. Andreeta and A. C. Hernandes**
*Grupo de Crescimento de Cristais e Materiais Cerâmicos, – MCDCM, IFSC - USP CEP 13560-970, P.O. Box: 369, São Carlos, SP, Brazil*

**E. R. Leite**
*Departamento de Química, LIEC-MCDCM, UFSCar, Caixa Postal 676, São Carlos SP, 13565-905 Brazil*

**O. F. de Lima**
*Instituto de Física Gleb Wataghin, UNICAMP, 13083-970 Campinas, SP, Brazil*

**A. W. Mombrú and R. Faccio**
*Laboratorio de Cristalografia y Química del Estado Sólido, Facultad de Química, Universidad de la República, P.O. Box 1157, Montevideo, Uruguay*

We have grown high-quality single crystals of the colossal-magnetoresistance (CMR) material $La_{0.7}Ca_{0.3}MnO_3$ by using the *laser heated pedestal growth* (LHPG) method. Samples were grown as fibers of different diameters, and with lengths of the order of centimeters. Their composition and structure were verified through X-ray diffraction, scanning electron microcopy with EDX *(Energy Dispersive X-ray Analysis)* and by Rietveld analysis. The quality of the crystalline fibers was confirmed by Laue and EBSD (*Electron Backscatter Diffraction*) patterns. Rocking curves performed along the fiber axis revealed a half-height width of 0.073º. The CMR behavior was confirmed by electrical resistivity and magnetization measurements as a function of temperature.

* Corresponding author; e-mail address: faraujo@df.ufscar.br

Magnetic manganites of the form $R_{1-x}A_xMnO_3$ [R = La, Pr, Nd; A = Ca, Sr, Ba] constitute one of the most fascinating classes of materials, exhibiting a wide variety of structures and properties. In particular, for x = 1/3, magnetic manganites exhibit colossal magnetoresistance (CMR) [1]. Their study has been specially renewed after the discovery of the CMR effect in thin films of these materials[2]. They are being pursued both for their potential applications[3] and for understanding the fundamental mechanisms governing their unique properties[4]. Some of their unusual properties have been tentatively explained by several models based on the Jahn-Teller distortion, double-exchange interaction, antiferromagnetic super-exchange, charge-orbital ordering interaction, and phase separation[5]. However, a complete understanding of the basic mechanisms involved is still lacking[4]. The central role played by compositional variations in these materials combined with inhomogeneities in its magnetic properties has been remarkable. Therefore, to advance the basic research about the magnetic manganites, in a reliable and meaningful way, it is of crucial relevance the use of high-quality single crystalline samples.

We report here on the growth and characterization of high-quality single crystals of $La_{0.7}Ca_{0.3}MnO_3$ (LCMO) by using the *Laser Heated Pedestal Growth* (LHPG) technique. Source pedestals of $La_{0.67}Ca_{0.33}MnO_3$ were initially prepared through a solid-state reaction route. Stoichiometric amounts of high purity $La_2O_3$, $CaCO_3$ and MnO were mixed and fired at 1200 ºC for 3 sessions of 24 hours each, with intermediate grindings and x-ray diffraction characterization to check the phase formation. The resulting powder was cold extruded with the same methodology described earlier[6]. The growth process was performed in a conventional LHPG system, with a $CO_2$ laser (Synrad, 60-1 - 125 W, CW)[6,7]. The fiber and pedestal pulling speeds, as well as the laser power, were modulated by an automatic diameter control system.[7] The average pulling speed was 18 mm/h. The pulling process of all pedestals was characterized by a stable behavior of the molten zone, in spite of an increase of the diameter of the pedestal.

The obtained fibers are single crystals with black surfaces. No inclusions of other phases were detected in their volume or surface. The fibers have a diameter around 1 mm and are up to 30 mm long. A longitudinally polished fiber was examined in a metallographic microscope under polarized light. From the seed, the fiber starts to grow as a set of long grains. After a few millimeters one grain orientation becomes preferred and the entire fiber becomes single-crystalline from this point on. The fiber composition was verified by EDX (*Energy Dispersive X-ray Analysis*) and X-rays diffraction (XRD) measurements. XRD patterns of grinded fibers (Fig. 1) present only peaks identified as $La_{0.7}Ca_{0.3}MnO_3$. The quality of the crystals was confirmed by Laue and EBSD (*Electron Backscatter Diffraction*) patterns and rocking curves performed along the fiber axis. This characterization provides a half-height width of 0.073º (inset, Fig. 1) besides showing that the fiber grows along the [100] direction. EDX analysis of the polished fiber detected only the expected LCMO phase. However, measurements performed at several points of the same fiber presented a slight and gradual increase of the calcium content from the seed region down to the solidification front. Also, the Mn/(La+Ca) ratio is less than unity, indicating the presence of vacancies in the Mn sub-lattice. In fact, the structure of LCMO allows the existence of cation vacancies in both La and Mn sub-lattices.

Then, the formulae of this compound may be expressed as $(LaCa)_{1-x}(\ )_xMn_{1-y}(\ )_yO_3$, where   represents a vacancy and the values of x and y can be as high as[8] 0.02. In spite of the presence of vacancies, the Mn content is constant along the entire fiber. The composition of the fiber varies, therefore, from $La_{0.77}Ca_{0.23}Mn_{0.98}O_3$ in the extremity near the seed, to $La_{0.71}Ca_{0.29}Mn_{0.98}O_3$ in the other extremity. This happens because the effective distribution coefficient for calcium is $K_{Ca} = C_s/C_m<1$, where $C_s$ and $C_m$ are the cation concentrations in the solid and melt, respectively. The observed value $K_{Ca} = 0.56$, obtained by direct comparison of the calcium concentration in both sides of the solidification front, is very close to the result $K_{Ca} = 0.57$, verified in crystals grown by floating zone method[9] (FZM). In fact, it was observed that the molten zone presents higher concentrations of both calcium and manganese, reaching Ca/La ratios as high as 1.37.

The fact that the Curie temperature $T_C$ in the LCMO system is strongly dependent on the $Mn^{3+}$-$Mn^{4+}$ ratio, which is determined by the calcium concentration, makes possible to correlate $K_x$ and $T_C$.

In order to do that, a single fiber was cut in five segments and the magnetization as a function of temperature was measured for each segment separately. The result is presented in Fig. 2. The segment nearest to the seed presented the lowest $T_C \approx 200$ K, which increased steadily up to $T_C \approx 270$ K at the other extremity. Comparing these values with the well-known magnetic phase diagram for LCMO[4], it can be inferred from these measurements that the calcium concentration raises from 0.2 to 0.35 as the fiber was pulled. Comparing the value near the seed (0.2), with the starting calcium content in the feed pedestal (0.33), one obtains a ratio of $K_{Ca} = 0.606$, which is consistent with the value $K_{Ca} = 0.56$ found by measuring the fiber composition directly.

As the solidification is carried on, there is an increase in the calcium content in the melt (due to the Ca segregation). However, as a significant fraction of the Ca ions remains in the molten zone, it is reasonable to expect a concomitant increase in the calcium content in the crystal. At a certain point, the amount of calcium going from the pedestal to the molten zone may become equal to the amount of calcium incorporated to the crystal. At this point the crystal became homogeneous with respect to calcium content. The results presented in Fig. 2 are consistent with this hypothesis: the content of calcium in the crystal increases with the concentration of Ca in the melt until an equilibrium situation is reached ($x \approx 0.35$).

Further investigation of the CMR behavior was performed by electrical resistivity measurements as a function of temperature. These results, obtained by using the conventional four-points method, are presented in Fig. 3. The crystal presents a semiconductor-like behavior at temperatures above $T_C$. Below this temperature the resistivity also changes dramatically to a metallic behavior. It is important to notice that the metallic behavior is observed in the as-grown crystals down to the lowest probed temperature of 4 K. That should be contrasted with the exponential increase of the resistivity at temperatures below 130 K reported for as-grown crystals obtained by the FZM method[10]. On the other hand, further annealing of the sample in oxygen atmosphere at 900º C and 1030 º C for 30 and 50 hours respectively, did not affect its resistivity behavior as a function of temperature. These results indicate that the oxygen content in the as-grown crystals is rather close to the ideal value, a quite surprising and important result. The colossal magnetoresistance, $\Delta R/R(H)$, for the as-grown crystal is presented in the inset of Fig. 3. It shows the expected behavior of LCMO samples as the external applied field is increased.

In summary, we report the growth of LCMO manganite single crystals by the LHPG technique. The effective distribution coefficient for calcium was found to be $K_{Ca} = 0.56$, in close agreement with the literature. The obtained crystals presented a small mosaicity, complete absence of inclusions, a sharp metal-insulator transition at the Curie temperature, with the metallic behavior subsisting down to 4 K, and a significant CMR of more than 250 % for H = 5 T. Our results indicate that the obtained crystals are of quality comparable or even higher than crystals obtained by other techniques[11]. Thus, we conclude that the method described here is suitable to grow CMR high-quality single crystals.

The authors gratefully acknowledge the financial support from Uruguayan agencies PEDECIBA and CSIC and Brazilian Agencies FAPESP, CAPES, PRONEX and CNPq.

**References**

[1] G. H. Jonker, and J. H. van Santen, Physica **16**, 337 (1950).

[2] R. von Helmolt, J. Wecker, B. Holzapfel, L. Schultz, and K. Samwer, Phys. Rev. Lett. **71**, 2331 (1993); S. Jin, T. H. Tiefel, M. McCormack, R. A. Fastnacht, R. Ramesh, and L. H. Chen, Science **264**, 413 (1994).

[3] M. Rajeswari, C. H. Chen, A. Goyal, C. Kwon, M. C. Robson, R. Ramesh, T. Venkatesan, and S. Lakeou, Appl. Phys. Lett. **68**, 3555 (1996).

[4] For a recent discussion and earlier references, see *Colossal Magnetoresistive Oxides*, ed. by Y. Tokura (Gordon and Breach, New York, 2000).

[5] C. Zener, Phys. Rev. B **82**, 403 (1951); P. G. de Gennes, Phys. Rev. **118**, 141 (1960); A. J. Millis, P. B. Littlewood, and B. I. Shraiman, Phys. Rev. Lett. **74**, 5144 (1995); G.-M. Zhao, K. Conder, H. Keller, and K.A. Muller, Nature **381**, 676 (1996).

[6] E. R. M. Andreeta, M. R. B. Andreeta, and A. C. Hernandes, J. Crystal Growth **234**, 782 (2002)

[7] M. R. B. Andreeta, L. C. Caraschi, and A. C. Hernandes, Materials Research v.6 (1), 107 (2003).

[8] B. C. Tofield, and W. R. Scott, Solid State Chem. **10**, 183 (1974).

[9] A. M. Balbashov, S. G. Karabashev, Y. M. Mukovskiy, and S. A. Zverkov, J. Crystal Growth **167**, 365 (1996).

[10] D. Shulyatev, S. Karabashev, A. Arsenov, and Y. Mukovskii, J. Crystal Growth **199**, 511 (1999).

[11] A. M. Balbashov, and S. K. Egorov, J. Crystal Growth **52**, 498 (1981); Y. Moritomo, Phys. Rev. B **51**, 16491 (1995).

**Figure captions**

**Figure 1 –** XRD patterns of grinded fibers, evidencing only peaks identified as $La_{0.7}Ca_{0.3}MnO_3$; the inset shows a rocking curve performed along the fiber axis providing half-height width of 0.073º.

**Figure 2 –** Magnetization vs. temperature curves (H = 0.1 T), for different segments of the fiber with different calcium concentrations, x, and $T_C$ values; the segment nearest to the seed shows the lowest $T_C \approx 200$ K, which increases steadily up to $T_C \approx 270$ K, at the other extremity.

**Figure 3 –** Electrical resistivity vs. temperature curves, for as-grown crystal with x ≈ 0.35 and different applied magnetic fields (indicated in the figure). The crystal shows a semiconductor-like behavior at temperatures above $T_C$, where it undergoes a metal-insulator transition.

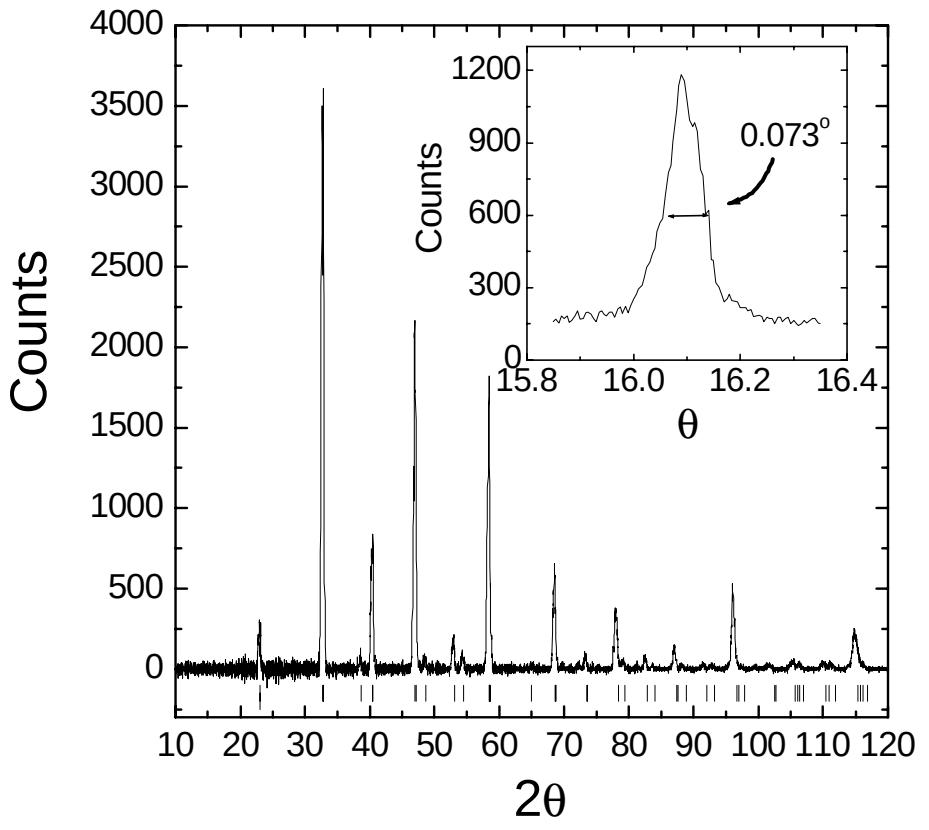

Figure 1 – C. Cardoso *et al*.

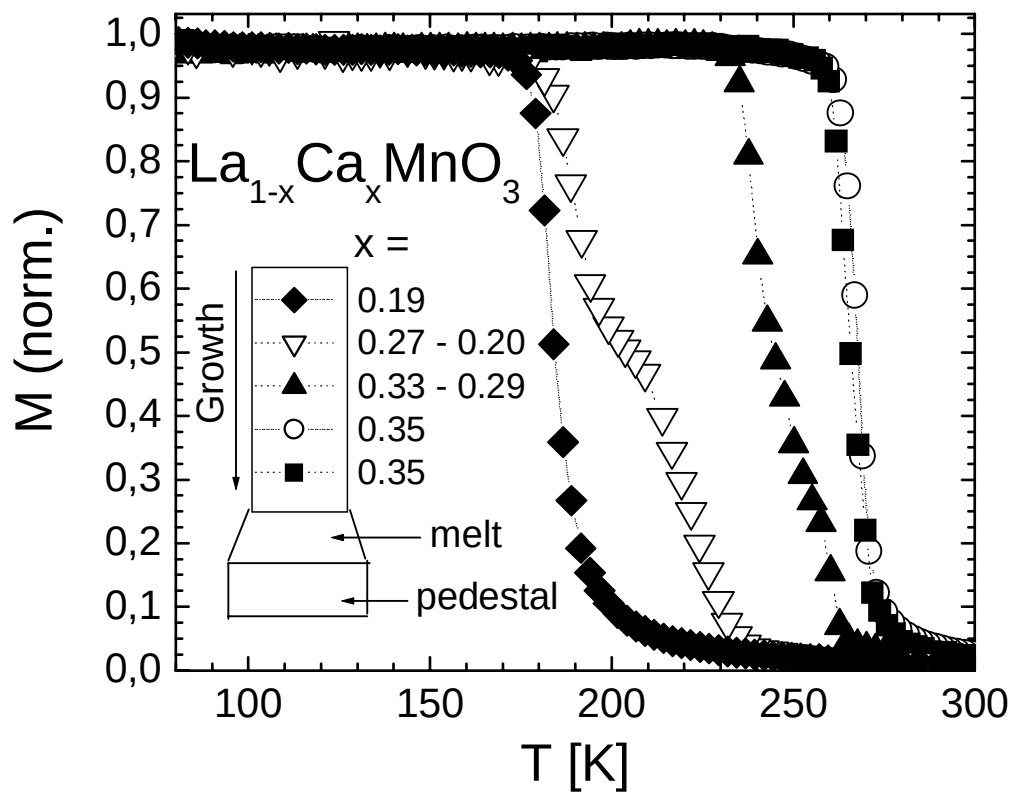

Figure 2 – C. Cardoso *et al.*

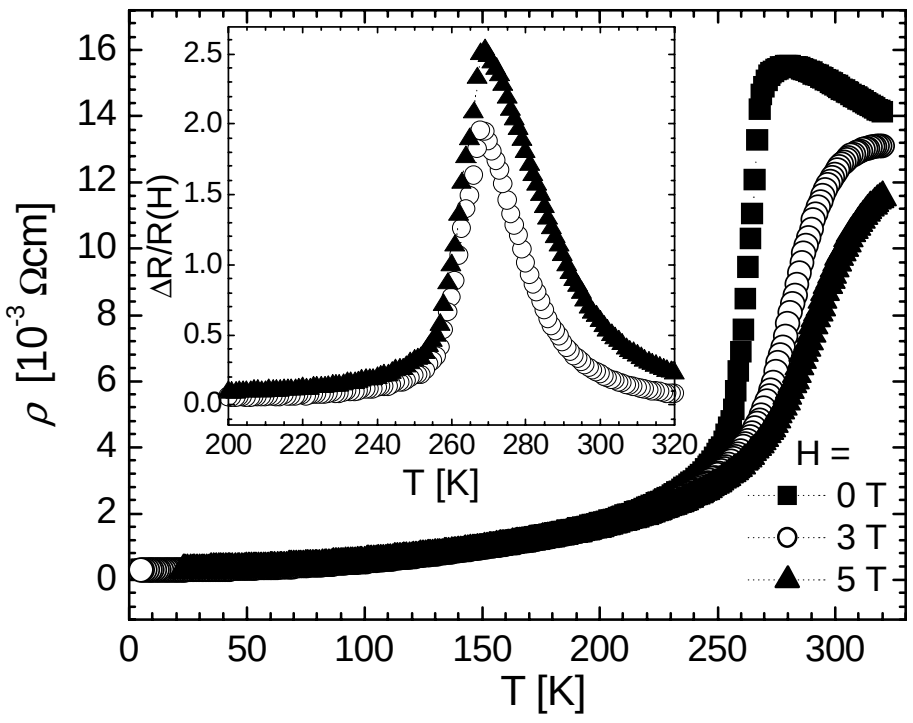

Figure 3 – C. Cardoso *et al.*